\documentclass[12pt]{article}
\usepackage[centertags]{amsmath}
\usepackage{amsfonts}
\usepackage{amssymb}
\usepackage{amsthm}
\newcommand{\beq}{\begin{equation}}
\newcommand{\eeq}[1]{\label{#1}\end{equation}}
\newcommand{\bea}{\begin{eqnarray}}
\newcommand{\eea}[1]{\label{#1}\end{eqnarray}}
\newcommand{\Dirac}{/\!\!\!\!D}
\newcommand{\Tr}{{\rm Tr}\,}
\begin{document}
\setlength{\topmargin}{-1cm} \setlength{\oddsidemargin}{0cm}
\setlength{\evensidemargin}{0cm}
\begin{titlepage}
\begin{center}
{\Large \bf Causal Propagation of a Charged Spin 3/2 Field in an External
Electromagnetic Background}

\vspace{20pt}

{\large Massimo Porrati and Rakibur Rahman\footnote{Address after October 1, 2009: Scuola Normale Superiore, Piazza dei Cavalieri 7, I-56126 Pisa (Italy).} }

\vspace{12pt}

Center for Cosmology and Particle Physics\\
Department of Physics\\ New York University\\
4 Washington Place\\ New York, NY 10003, USA

\end{center}
\vspace{20pt}

\begin{abstract}
We present a  Lagrangian for a massive,
charged spin 3/2 field in a constant external
electromagnetic background, which correctly propagates only
physical degrees of freedom inside the light cone. The Velo-Zwanziger
acausality and other pathologies such as loss of hyperbolicity or the
appearance of unphysical degrees of freedom are avoided by a
judicious choice of non-minimal couplings. No additional fields or equations
besides the spin 3/2 ones are needed to solve the problem.
\end{abstract}

\end{titlepage}

\newpage
While completely explicit actions of {\em free}
massive fields of spin arbitrarily
larger than one $-$ which propagate within the
light cone the correct number of physical degrees of freedom $-$
have been known since the 1970's~\cite{sh}, consistent actions
for interacting fields have been much hard to construct. Indeed, even the
conceptually simpler problem of describing high-spin particles in fixed
external field backgrounds
has proved itself fraught with difficulties. As already noticed by
Fierz and Pauli
seventy years ago~\cite{fp}, a Lagrangian formulation of interacting high-spin
fields is essential even at the classical level, to avoid algebraic
inconsistencies in the equations of motion. However, when the high-spin field
is coupled to either external or dynamical fields, a Lagrangian formulation
guarantees neither that no unphysical degrees of freedom start propagating, nor
that the physical ones propagate only causally.

This pathology is particularly vexing for the seemingly simple case
of charged, massive particles of spin 3/2. Their well-known free
action was found in 1941 in~\cite{rs}, but it took many years before
realizing that minimal coupling to external electromagnetic fields
resulted in equations of motion which exhibited faster-than-light
propagation of signals~\cite{vz} (see also~\cite{hor}). This lack of
causality also shows up in higher spin fields, such as spin
2~\cite{vks}.\footnote{The causality problem is a classical
pathology; its quantum analog is that canonical commutators become
ill-defined. The latter was noticed in~\cite{js} long
before~\cite{vz}.}

Massive, electrically charged states of spin 3/2 or higher do exist
in QCD as resonances. Moreover, open string theory contains
(infinitely) many charged, massive particles of spin higher than
one.~\footnote{Light spin 3/2 particles may also appear in
Randall-Sundrum models~\cite{hmrr}.} Both string theory and QCD are
to the best of our understanding consistent and causal, especially
in the dynamical regime describing particles in fixed external
electromagnetic fields. So, a natural question to ask is how the
Velo-Zwanziger acausality problem is resolved, first of all in the
simplest setting of them all: spin 3/2.
\subsection*{Possible Solutions}
Various scenarios exist for rescuing causality.
\subsubsection*{Adding New Degrees of Freedom}
One is that a single charged spin 3/2 field is
inconsistent or non-causal when considered in isolation. It could happen that
causality forces upon us the existence of other fields besides the spin 3/2
one.
After all, we do know an example of consistently propagating charged spin
3/2 fields: $\mathcal{N}=2$ extended ``gauged'' supergravity~\cite{fpvn}.
In $\mathcal{N}=2$ theories,
the gravitino can be charged under a $U(1)$ field (the graviphoton).
Supersymmetry can be broken without introducing a cosmological
constant~\cite{dwvp,ss}, resulting in a massive spin 3/2 field propagating
in flat space. Causality in this case is due to gravitational back-reaction.

More specifically, as shown in~\cite{dpw}, superluminal propagation of the
mass $m$, charge $e$ gravitino would occur in flat space when, in some
Lorentz frame, the magnetic field $\bf B$ attains the critical value
$|{\bf B}|=\sqrt{3}m^2/e$.
In that frame, the energy density
$T_{00} = \tfrac{1}{2}({\bf E}^2 + {\bf B}^2)$ is
always larger than $\tfrac{3}{2}m^4/e^2$. Since in this theory
gravity is dynamical, the
gravitational back-reaction induces a curvature in space-time,
characterized by a length scale
$L^{-2} ={\cal O}(3m^4/2M_{\text{Pl}}^2 e^2)$. But in $\mathcal{N}=2$
theories, the graviphoton
charge of the gravitino and the Planck mass $M_{\text{Pl}}$ are related by
$e= m/M_{\text{Pl}}$; therefore, space-time is significantly curved already
at the Compton wavelength scale of
the gravitino $L =1/m$. This is precisely
the regime where the flat-space causality results of~\cite{vz,dpw}
cease to apply. Indeed, ref.~\cite{dw} extended the causality analysis
done for pure supergravity in~\cite{dz} to prove that $\mathcal{N}=2$
supergravity is causal and hyperbolic when
$m>\sqrt{\tfrac{2}{3}}e M_{\text{Pl}}$.~\footnote{Partial results on causality
of ${\cal N}=2$ and Kaluza-Klein supergravities
can also be found in~\cite{rsi}.}

The main drawback of extended supergravity is that it cannot solve the
causality problem of spin 3/2 fields unless the charge obeys the
``Kaluza-Klein'' relation $e=m/M_{\text{Pl}}$. When $-$ as for electromagnetic
interactions $-$ $e$ is fixed ($e\approx 0.3$),
the gravitational back-reaction of spin 3/2 particles much lighter than
${\cal O}(eM_{\text{Pl}})$ is negligible, so they
can still propagate superluminally.

\subsubsection*{Adding Non-Minimal Terms}

A different solution to the causality problem may be to change the minimal
spin 3/2 theory not by adding new dynamical degrees of freedom, but simply
by adding non-minimal gauge invariant interactions. That this could be the right solution
is strongly suggested by analogy with the
only known example of a consistent model of high-spin particles
of arbitrary charge, which propagates causally in an external
electromagnetic field, constant but otherwise arbitrary.
This is the Argyres-Nappi action~\cite{an}. It describes a single, massive
spin 2 field, charged under a $U(1)$. Charge and mass are independent
variables; in particular, a dynamical regime exists which decouples
gravitational interactions, while keeping the $U(1)$ charge finite.
The Argyres-Nappi action is highly non-minimal: it is
quadratic in the charged spin 2 field but non-polynomial in the electromagnetic
field strength $F_{\mu\nu}$. It was obtained from the equations of motion of
charged open strings in a background with a nonzero, constant external
field strength $F_{\mu\nu}$.

Even though derived within string theory, the reason why the
Argyres-Nappi theory is causal and consistent is simple: After a
straightforward redefinition of variables, its equations
of motion enforce the standard transverse-tracelessness constraint on the
spin 2 field $h_{\mu\nu}$. By substituting the constraint into the equations
of motion, one obtains a good hyperbolic system,
$\square h_{\mu\nu} + \mbox{ lower derivative terms} =0$,
which manifestly propagates
five degrees of freedom within the light cone.

It would be odd if what works with spin 2 does not work with spin 3/2,
especially since the reason for causality in the Argyres-Nappi action is
not due to exotic properties of string theory, but rather to a clever
combination of non-minimal terms. So, even for spin 3/2,
it makes sense to consider a general non-minimal Lagrangian of the
form~\footnote{Our conventions are as follows: the metric $\eta_{\mu\nu}$
is mostly plus, $\bar{\psi}_\mu = \psi^\dagger_\mu \gamma^0$,
$\gamma^{\mu\,\dagger}=\eta^{\mu\mu}\gamma^\mu$,
$\gamma^5=-i\gamma^0\gamma^1\gamma^2\gamma^3$. We always antisymmetrize
with unit strength:
$\gamma^{\mu_1....\mu_n}= \tfrac{1}{n!}\gamma^{\mu_1}\gamma^{\mu_2}..\gamma^{\mu_n}
+ \mbox{ antisymmetrization}$.}
\bea
L &=& -i\bar{\psi}_\mu A^{\mu\nu\rho}(F)D_\nu \psi_\rho -
i\bar{\psi}_\mu B^{\mu\nu}(F)\psi_\nu, \nonumber \\
A^{\mu\nu\rho}(F)&=&\gamma^{\mu\nu\rho} + {\cal O}(F), \qquad
B^{\mu\nu}(F)= m\gamma^{\mu\nu} + {\cal O}(F).
\eea{m1}
The non-minimal couplings $A^{\mu\nu\rho}(F),B^{\mu\nu}(F)$ are
functions of the electromagnetic field
strength $F_{\mu\nu}$, analytic near $F_{\mu\nu}=0$.
Their form will be specified later.
\subsubsection*{What We Cannot Expect to Find}
Before analyzing further eq.~(\ref{m1}), it is important to understand clearly
what problem we must solve and which one we should not.
Our aim is to find a Lagrangian that propagates within the light cone
only four degrees of freedom $-$ the four physical helicities
of a spin 3/2 field $-$
in an external electromagnetic background. Our method will work for constant
backgrounds. While this is a drawback, it does take care of the original
Velo-Zwanziger problem, which manifests itself already
for constant backgrounds~\cite{vz}.

We do not want to find a Lagrangian that works for arbitrarily large values
of the relativistic field invariants $F_{\mu\nu}F^{\mu\nu}$,
$F_{\mu\nu}\tilde{F}^{\mu\nu}$~\footnote{$\tilde{F}_{\mu\nu}=\tfrac{1}{2}
\epsilon_{\mu\nu\rho\sigma}F^{\rho\sigma}$ with $\epsilon_{\mu\nu\rho\sigma}$
totally antisymmetric and normalized as $\epsilon_{0123}=+1$.}. The reason is
that whenever these invariants become ${\cal O}(m^4/e^2)$,
several instabilities
appear, that make the very concept of a long-lived, propagating spin 3/2 field
unphysical. One such instability is the Schwinger pair production~\cite{s},
which
becomes significant when
$F_{\mu\nu}\tilde{F}^{\mu\nu}=0$ and $F_{\mu\nu}F^{\mu\nu}\sim -m^4/e^2$.
Another is the spin 3/2 analog~\cite{fpo}
of the Nielsen-Olesen instabilities~\cite{no}, which appear when
$F_{\mu\nu}\tilde{F}^{\mu\nu}=0$ and $F_{\mu\nu}F^{\mu\nu}\sim +m^4/e^2$.
Though these instabilities are normally said to occur when either the
electric field (Schwinger) or the magnetic field (Nielsen-Olesen) are
${\cal O}(m^2/e)$, it is important to realize that they only depend on
relativistic invariant combinations of the field strength. These
instabilities mean
that whatever Lagrangian one may use to describe a spin 3/2 field in isolation,
it will always be only an effective one, reliable only when energies are
sufficiently small {\em and the
relativistic field invariants are much smaller than} ${\cal O}(m^4/e^2)$.
It is thus particularly telling that the Argyres-Nappi Lagrangian
becomes ill-defined precisely when the relativistic field invariants reach
their critical strength $\sim m^4/e^2$~\cite{an}.

The Velo-Zwanziger problem is different in that it persists even at
arbitrarily small values of the relativistic field invariants.
Concretely, in the minimal model, the
magnetic field ${\bf B}$ can reach its critical value
$|{\bf B}|=\sqrt{\tfrac{3}{2}}m^2/e$
~\footnote{For minimal supergravity the critical value instead
is $|{\bf B}|=\sqrt{3}m^2/e$.}
in a frame where $|{\bf E}|=\sqrt{\tfrac{3}{2}}m^2/e -\epsilon$,
with $\epsilon$ an arbitrarily small
number. So, it is a real problem that occurs within the regime of validity of
the effective theory.

This is the problem we need to solve: we must find a non-minimal
Lagrangian that
propagates causally the correct number of degrees of freedom whenever
$|F_{\mu\nu}F^{\mu\nu}| \ll m^4/e^2$,
$|F_{\mu\nu}\tilde{F}^{\mu\nu}| \ll m^4/e^2$.

We shall not worry if the Lagrangian fails whenever either of these invariants
becomes ${\cal O}(m^4/e^2)$, since in any case {\em any} Lagrangian treating
the electromagnetic background as fixed  is meaningless, because it
fails to take into account large effects due to {\em electromagnetic}
back-reaction.
\subsection*{What is Not a Solution}
Hermiticity of the Lagrangian in eq.~(\ref{m1}) imposes some constraints on the
coefficients $A^{\mu\nu\rho}(F),B^{\mu\nu}(F)$, namely
\beq
\gamma^0 (A^{\rho\nu\mu})^{\dagger} \gamma^0= A^{\mu\nu\rho}, \qquad
\gamma^0 (B^{\nu\mu})^{\dagger} \gamma^0 = -B^{\mu\nu}.
\eeq{m2}
Moreover, unless $A^{\mu\nu\rho}$ and $B^{\mu\nu}$ are antisymmetric in their
vector Lorentz indices, eq.~(\ref{m1}) propagates additional degrees of
freedom whenever $F_{\mu\nu}\neq 0$. These degrees of freedom are dangerous
because they interact through relevant {\em and} irrelevant interactions with
the electromagnetic field, but are absent at $F_{\mu\nu}=0$,
when the coefficients in the
Lagrangian~(\ref{m1}) are antisymmetric. So, their kinetic term is proportional
to $|F_{\mu\nu}|$ and thus the strength of all their irrelevant interactions
diverges in the weak field limit $|F_{\mu\nu}|\rightarrow 0$. The existence of
these unwanted degrees of freedom makes the solution of the Velo-Zwanziger
problem proposed in the appendix of~\cite{fpt} unacceptable, as pointed out
in~\cite{dpw}. Yet, the idea that non-minimal interactions may
cure the problem can be salvaged from that work, as we shall proceed to
explain.
\subsection*{Constraints}
What makes the Argyres-Nappi action work is that on an appropriately redefined
spin 2 field it enforces the {\em same} constraint as the free action does,
namely transverse-tracelessness~\footnote{In the notation of ref.~\cite{an} the
transverse-traceless field is $(HhH^*)_{\mu\nu}$.}.
Similarly, here we demand that our non-minimal action enforce the constraint
\beq
\gamma^\mu \psi_\mu =0.
\eeq{m3}
We will present later a non-minimal action satisfying this requirement. It
will turn out to
have a canonical kinetic term $A^{\mu\nu\rho}=\gamma^{\mu\nu\rho}$.
Before entering into the details of its construction,
it is instructive to see why
constraint~(\ref{m3}) ensures at once that the equations of motion derived from
Lagrangian~(\ref{m1}) define a hyperbolic
system that propagates causally four degrees of freedom.

When $A^{\mu\nu\rho}=\gamma^{\mu\nu\rho}$, the equations of motion are
\beq
R^\mu  + B^{\mu\nu}\psi_\nu =0, \qquad
R^\mu\equiv \gamma^{\mu\nu\rho}D_\nu \psi_\rho,
\eeq{m4}
while their gamma-trace is
\beq
2\gamma^{\mu\nu}D_\mu \psi_\nu + .... =0
\eeq{m5}
where the ellipsis stand for ``mass'' terms containing no derivatives.
Using the identity
\beq
\gamma^{\mu\nu\rho} =\gamma^\mu\gamma^{\nu\rho} - \eta^{\mu\nu} \gamma^\rho
+ \eta^{\mu\rho} \gamma^\nu ,
\eeq{m6}
eq.~(\ref{m5}) and the constraint~(\ref{m3}), one can reduce
equations of motion~(\ref{m4}) to a standard, manifestly causal Dirac form:
\beq
\Dirac \psi_\mu + \mbox{ non-derivative terms} =0.
\eeq{m7}
Since $B^{\mu\nu}$ is  antisymmetric in $\mu,\nu$, the 0-th component of the
equations of motion, $R^0=0$  contains neither time derivatives nor $\psi_0$;
thus, it enforces four constraints among the remaining fields
$\psi_i$, $i=1,2,3$. Constraint~(\ref{m3}) then removes
$\psi_0=\gamma^0\gamma^i \psi_i$ leaving $3\times 4 - 4 = 8$ physical
variables i.e. four degrees of freedom (four coordinates and four conjugate
momenta). A completely analogous way to prove the same result
uses the obvious fact that consistent propagation of the
constraint~(\ref{m3}) and eq.~(\ref{m7}) imply $D^\mu \psi_\mu =$
non-derivative terms; so, using eq.~(\ref{m7}) to  write
$D_0\psi_0 = \gamma^0 \gamma^i D_i \psi_0$ + non-derivative terms, one gets
from the above divergence the four additional constraints needed to reduce the
number of degrees of freedom to four.

\subsection*{Construction of the Non-Minimal Action}
Of course, the real question is whether a non-minimal action that gives
eq.~(\ref{m3}) exists. We prove that it does by explicitly constructing one.

Our ansatz for the non-minimal ``Pauli'' terms is
\bea
A_{\mu\nu\rho}&=&\gamma_{\mu\nu\rho}, \label{m8}\\
B_{\mu\nu} &=& m\gamma_{\mu\nu}+G^+_{\mu\nu} + \gamma^\rho T_{\rho[\mu}\gamma_{\nu]},
\label{m9} \\
G^+_{\mu\nu} &\equiv & G_{\mu\nu} +
\tfrac{1}{2}\gamma_{\mu\nu\rho\sigma}G^{\rho\sigma}
\eea{m10}
The Lorentz tensor $G_{\mu\nu}$ is antisymmetric
($G_{\mu\nu}=-G_{\nu\mu}$) and ${\cal O}(F)$, while the Lorentz tensor $T_{\mu\nu}$ is
symmetric and traceless ($T_{\mu\nu}=T_{\nu\mu}$, $T_\mu^\mu=0$)
and ${\cal O}(F^2)$.
Hermiticity of Lagrangian~(\ref{m1}) implies that $T_{\mu\nu}$ is real and
$G_{\mu\nu}$ is imaginary. Apart from these constraints, they are as-yet
unspecified
functions of the electromagnetic field strength $F_{\mu\nu}$.

As pointed out in ref~\cite{dpw}, addition of $G^+_{\mu\nu}$ alone
can never yield a causal theory, irrespective of its functional
dependence on $F_{\mu\nu}$. It is crucial to notice that the term
proportional to $T_{\mu\nu}$ instead, is structurally different from
all those studied in~\cite{dpw}~\footnote{Appendix B of~\cite{dpw}
proves that non-minimal terms of the generic form $B_{\mu\nu} =
iW_{\mu\nu} + \gamma^5 X_{\mu\nu} + \gamma_{\mu\nu}Y + i\gamma^5
\gamma_{\mu\nu} Z$ still allow for superluminal propagation, even
when the coefficients $W,X,Y,Z$ are arbitrary functions of
$F_{\mu\nu}$. To the best of our knowledge, the last term in our
eq.~(\ref{m9}) has never been considered before.}.

A few identities that will be crucial for our construction and
follow from elementary manipulations of gamma-matrix
algebra are~(\ref{m6}) and
\bea
\gamma^\mu G^+_{\mu\nu}&=& \tfrac{1}{2} \gamma \cdot G \gamma_\nu, \qquad
\gamma \cdot G \equiv \gamma_{\rho\sigma} G^{\rho\sigma}, \label{m11} \\
G^+_{\mu\nu} &=& -\tfrac{1}{4} \gamma^\rho \gamma \cdot G \gamma_{\rho\mu\nu}
\label{m12}\\
\gamma^\rho D_{[\rho} \psi_{\mu]} &=&\tfrac{1}{2} R_\mu -\tfrac{1}{4} \gamma_\mu
\gamma^\rho  R_\rho , \qquad \gamma^{\mu\nu}D_\mu \psi_\nu =
\tfrac{1}{2} \gamma^\rho R_\rho.
\eea{m12a}
Either direct calculation or simple considerations of representation theory of
the Lorentz group lead to another important identity~\footnote{$G_{\mu\nu}$
decomposes into irreps of $SL(2,C)$ as $(1,0) +(0,1)$ while
$\tilde{G}_{\mu\nu}$ decomposes as $(1,0) - (0,1)$. The most general tensor
product of the two decomposes as $(2,0) + (1,0) + (0,1) + (0,0)$, but since
its antisymmetric part
vanishes on self-dual or anti self-dual backgrounds, it cannot
contain either $(1,0)$ or $(0,1)$. On the other hand, the same tensor product
can only contain representations appearing in the tensor product
$(\tfrac{1}{2},\tfrac{1}{2})\times (\tfrac{1}{2},\tfrac{1}{2})=
(1,1) + (1,0) + (0,1) + (0,0)$. The only common element is $(0,0)$.}
\beq
G_{\mu\rho}\tilde{G}^{\rho\nu} = -\tfrac{1}{4} \delta_\mu^\nu
G_{\rho\sigma}\tilde{G}^{\rho\sigma}.
\eeq{m13}

Thanks to these identities, the gamma-trace of the equations of motion~(\ref{m4})
is
\beq
2\gamma^{\mu\nu} D_\mu \psi_\nu + T^{\mu\nu}\gamma_\mu\psi_\nu =
[-3m + {\cal O}(F)]\,\gamma\cdot \psi .
\eeq{m14}
The term multiplying $\gamma\cdot \psi\equiv \gamma^\mu \psi_\mu$ on the
right-hand side of this equation is a $4\times 4$
matrix containing no derivatives, thus
built only out of gamma matrices, $G_{\mu\nu}$, and $T_{\mu\nu}$. The split
into the constant term $-3m$ and higher powers of the electromagnetic field
follows simply from our ansatz, $G_{\mu\nu}={\cal O}(F)$,
$T_{\mu\nu}={\cal O}(F^2)$.

Next we take the divergence of equations of motion~(\ref{m4}).
Since the covariant derivative
$D_\mu$ obeys $[D_\mu,D_\nu]= ie F_{\mu\nu}$ we have
\beq
D_\mu R^\mu = -ie F^{\mu\nu} \gamma_\mu \psi_\nu + {ie\over 2}
\gamma \cdot F \gamma \cdot \psi.
\eeq{m15}

By using eq.~(\ref{m14}), identities~(\ref{m11}-\ref{m12a})
plus the vanishing of
$\gamma^\mu \gamma_{\alpha\beta} \gamma_\mu$ and
$\gamma^\mu T_{\mu\nu} \gamma^\nu$~\footnote{The first is a standard
gamma-matrix identity, while the second follows from tracelessness of $T_{\mu\nu}$.}
we  re-write the divergence as
\bea
&&
-ie\gamma_\mu F^{\mu\nu} \psi_\nu -\tfrac{1}{2}m\gamma_\mu T^{\mu\nu} \psi_\nu
-\tfrac{1}{4} \gamma^\mu \gamma \cdot G\,[\,m\eta_{\mu\nu} - G^+_{\mu\nu}
+ T_{\mu\nu}\,]\,\psi^\nu \nonumber \\ &&
-\tfrac{1}{2} \gamma_\rho T^{\rho\mu}\,[\, m\eta_{\mu\nu} -G^+_{\mu\nu} +
T_{\mu\nu}\,]\,\psi^\nu = \left[\,\tfrac{3}{2}m^2 + {\cal O}(F)\,\right]
\gamma \cdot \psi .
\eea{m16}
In this equation we use again identities~(\ref{m11}-\ref{m12a}) as well as
identity~(\ref{m13}) to simplify the term quadratic in $G$; we obtain
\bea && \gamma^\mu \left(-ieF_{\mu\nu} + mG_{\mu\nu}+ \tfrac{1}{2}
T_\mu^\rho G^+_{\rho\nu} - \tfrac{1}{2}G^+_{\mu\rho} T^\rho_\nu + G_{\mu\rho}
T^\rho_\nu\right)\,\psi^\nu \nonumber\\ &&
-\gamma^\mu\left(G_\mu^{\;\;\,\rho} G_{\rho\nu} + mT_{\mu\nu}
+\tfrac{1}{2}T_\mu^\rho T_{\rho\nu}\right)\,\psi^\nu=
\left[\,\tfrac{3}{2}m^2 + {\cal O}(F)\,\right] \gamma \cdot \psi .
\eea{m17}

Two conditions must be met to enforce the standard constraint
$\gamma \cdot \psi =0$. The first is that the left-hand side of
eq.~(\ref{m17}) must either vanish or be proportional to ${\cal O}(F)
\gamma \cdot \psi$; the second is that the matrix
$[\,\tfrac{3}{2}m^2 + {\cal O}(F)\,]$ is invertible.

The hard one is the first.

To satisfy it, we first of all set
\beq
T_\mu^\nu=
A\left( G_{\mu\rho}G^{\rho\nu} -\tfrac{1}{4} G_{\sigma\rho}G^{\rho\sigma}
\delta_\mu^\nu\right),
\eeq{m18}
where $A$ is a constant.
This choice renders the term
$T_\mu^\rho T_{\rho\nu}={\cal O}(F^4)\gamma\cdot\psi$ and also makes the term inside the first parenthesis in eq.~(\ref{m17}) antisymmetric in $\mu,\nu$. Both properties follow from eq.~(\ref{m13}), which gives the identities
($\Tr H \equiv H_\mu^\mu$)
\bea
G^+_{\mu\rho}G^{-\, \rho\nu}&=& G^-_{\mu\rho}G^{+\, \rho\nu}=
2\left[ G_{\mu\rho}G^{\rho\nu} -\tfrac{1}{4} \Tr (G^2) \delta_\mu^\nu\right]
\label{m19} \\
G^\pm_{\mu\rho}G^{\pm\, \rho\nu}&=&\tfrac{1}{2}[\,\Tr (G^2) \pm
i\gamma^5 \Tr (G\tilde{G})\,]\,\delta_\mu^\nu.
\eea{m20}
Now the two terms in parentheses in eq.~(\ref{m17}) must separately
either vanish or be proportional to $\gamma \cdot \psi$, since the first is
antisymmetric in $\mu,\nu$ while the second is symmetric.

The choice $A=-m^{-1}$ makes the whole symmetric term in~(\ref{m17}) equal to
${\cal O}(F^2)\gamma \cdot \psi$. On the other hand, the antisymmetric term
vanishes if $G_{\mu\nu}$ satisfies the following implicit equation:
\beq
G_{\mu\nu}= +{ie\over m} F_{\mu\nu} +{1\over 4m^2}\Tr (G^2) G_{\mu\nu} -
{1\over 4m^2}\Tr (G\tilde{G}) \tilde{G}_{\mu\nu}.
\eeq{m21}
This can be solved by power series as long as the relativistic field
invariants $F_{\mu\nu}F^{\mu\nu}$, $F_{\mu\nu}\tilde{F}^{\mu\nu}$ have magnitudes
that are small compared to $m^4/e^2$~\footnote{(Semi)-explicitly,
$G_{\mu\nu}=a F_{\mu\nu} + b \tilde{F}_{\mu\nu}$, with $a,b$ analytic
functions of the relativistic field invariants. They obey
$a=ie/m + {\cal O}[\Tr(F^2),\Tr(F\tilde{F})]$,
$b={\cal O}[\Tr(F^2),\Tr(F\tilde{F})]$ for $|\Tr(F^2)|, |\Tr(F\tilde{F})|\ll m^4/e^2$.}.
This is the crucial feature we need, namely a theory that
only breaks down for large {\em invariants}, but that is well-behaved when
they are  small, even when some field strength component becomes
${\cal O}(m^2/e)$.

Eqs.~(\ref{m18},\ref{m21}) make the constraint~(\ref{m17}) take the desired
form
\beq
\left[\,\tfrac{3}{2}m^2 + {\cal O}(F)\,\right] \gamma \cdot \psi =0.
\eeq{m22}
The proportionality matrix multiplying $\gamma \cdot \psi$ contains only
gamma matrices and powers of $(e/m^2)F_{\mu\nu}$ with dimensionless
coefficients; therefore, Lorentz invariance
implies that its determinant can only be a function of relativistic field invariants,
hence invertible when
$|F_{\mu\nu}F^{\mu\nu}|, |F_{\mu\nu}\tilde{F}^{\mu\nu}|\ll m^4/e^2$. So the
second condition is met precisely when an effective Lagrangian description is
supposed to make sense.

The redefinition $G_{\mu\nu} = im X_{\mu\nu}$ and a
straightforward computation give
the explicit form of the matrix; the constraint equation then becomes
\beq
 \left\{ 48 -
[\Tr(X^2)]^2 -  [\Tr(X \tilde{X})]^2 \right\} \gamma \cdot \psi =0,
\eeq{m23}
which manifestly depends only  on relativistic field invariants.
\subsection*{Summary}
The construction presented in this paper answers a question
that in various guises remained unanswered for many decades, namely: does a
consistent, causal Lagrangian  describing  a single
massive, charged particle of spin larger than one in interaction with the
electromagnetic field exist?

The answer for spin 3/2 is yes, at least for constant external fields.
This is a major achievement in itself, since constant fields are exactly
those that cause the Velo-Zwanziger acausality~\cite{vz} and the
Johnson-Sudarshan problem~\cite{js}.

The crucial property of our construction is that the standard
gamma-tracelessness constraint $\gamma \cdot \psi =0$ is enforced
{\em exactly}. Enforcing it only up to a finite order in an expansion in
powers of the field strength  would not suffice. To see this, we may try to
substitute $G_{\mu\nu}=i(e/m)F_{\mu\nu},T_{\mu\nu}=0$ in our non-minimal
ansatz eq.~(\ref{m9}). This choice satisfies constraint~(\ref{m3}) up to
${\cal O}(F^2)$:
\beq
\left[ \tfrac{3}{2} m^2 + {\cal O}(F)\right]\gamma\cdot \psi=
{e^2\over m^2}\gamma^\mu F_{\mu\rho}F^{\rho\nu}\psi_\nu .
\eeq{m24}
As shown in~\cite{dpw}, superluminal propagation of signals occurs when
$\psi_0 \neq 0$, $\psi_i=0$ solve this equation. Contrary to, say,
constraint
eq.~(\ref{m23}), eq.~(\ref{m24}) depends on quantities, such as the
electromagnetic stress energy tensor, that can be large even when the
relativistic field invariants are small. This property allows
$\psi_0 \neq 0$, $\psi_i=0$ to be a solution even when
$|\Tr(F^2)|, |\Tr(F\tilde{F})|\ll m^4/e^2$, as it can be easily
proven by direct computation~\cite{dpw}.

It is also illuminating that the solution involves no extra degrees of freedom
and that carefully chosen parity-preserving non-minimal terms suffice.
It is an amusing and perhaps deep
fact that the non-minimal couplings also give a gyromagnetic factor $g=2$ $-$ the
same value needed to improve the high-energy behavior of ``Compton'' forward
scattering amplitudes, and the one given by open string theory~\cite{fpt}.

It is finally worth noticing that the non-minimal terms required by causality
in our admittedly non-unique Lagrangian
lower the intrinsic UV cutoff of the theory, from its theoretical maximum
$\Lambda \sim e^{-1/2}m$~\cite{pr}, to $\Lambda \sim e^{-1/3}m$. If this
property were to extend to the most general causal Lagrangian of charged
spin 3/2 fields it would offer a powerful tool to establish stronger, model
independent limits on the UV cutoff of such theories.
\subsection*{Acknowledgments}
We would like to thank A. Waldron, especially for pointing out to us
ref.~\cite{dw}.
M.P. is supported in part by NSF grant  PHY-0758032, and
by ERC Advanced Investigator Grant n.226455
{\em Supersymmetry, Quantum Gravity and Gauge Fields (Superfields)}


\begin{thebibliography}{99}
\bibitem{sh}
  L.P.S.~Singh and C.R.~Hagen,
  Phys.\ Rev.\  D {\bf 9}, 898 (1974);
  Phys.\ Rev.\  D {\bf 9}, 910 (1974).
\bibitem{fp}
  M.~Fierz and W.~Pauli,
  Proc.\ Roy.\ Soc.\ Lond.\  A {\bf 173}, 211 (1939);
  Helv.\ Phys.\ Acta {\bf 12}, 297 (1939).
\bibitem{rs}
  W.~Rarita and J.~Schwinger,
  Phys.\ Rev.\  {\bf 60}, 61 (1941).
\bibitem{vz}
  G.~Velo and D.~Zwanziger,
  Phys.\ Rev.\  {\bf 186}, 1337 (1969);
  \bibitem{hor}
  M.~Hortacsu,
  Phys.\ Rev.\  D {\bf 9}, 928 (1974).
\bibitem{vks}
  G.~Velo,
  Nucl.\ Phys.\  B {\bf 43}, 389 (1972);
  M.~Kobayashi and A.~Shamaly,
  Phys.\ Rev.\  D {\bf 17}, 2179 (1978);
  Prog.\ Theor.\ Phys.\  {\bf 61}, 656 (1979).
\bibitem{js}
  K.~Johnson and E.~C.~G.~Sudarshan,
  Annals Phys.\  {\bf 13}, 126 (1961).
\bibitem{hmrr}
  B.~Hassanain, J.~March-Russell and J.~G.~Rosa,
  arXiv:0904.4108 [hep-ph].
\bibitem{fpvn}
  S.~Ferrara and P.~van Nieuwenhuizen,
  Phys.\ Rev.\ Lett.\  {\bf 37}, 1669 (1976).
\bibitem{dwvp}
  B.~de Wit, P.~G.~Lauwers and A.~Van Proeyen,
  Nucl.\ Phys.\  B {\bf 255}, 569 (1985).
\bibitem{ss}
  J.~Scherk and J.~H.~Schwarz,
  Phys.\ Lett.\  B {\bf 82}, 60 (1979);
  Nucl.\ Phys.\  B {\bf 153}, 61 (1979).
\bibitem{dpw}
  S.~Deser, V.~Pascalutsa and A.~Waldron,
  Phys.\ Rev.\  D {\bf 62}, 105031 (2000)
  [arXiv:hep-th/0003011].
\bibitem{dw}
  S.~Deser and A.~Waldron,
  Nucl.\ Phys.\  B {\bf 631}, 369 (2002)
  [arXiv:hep-th/0112182].
\bibitem{dz}
  S.~Deser and B.~Zumino,
  Phys.\ Rev.\ Lett.\  {\bf 38}, 1433 (1977).
\bibitem{rsi}
  S.~D.~Rindani and M.~Sivakumar,
  Phys.\ Rev.\  D {\bf 37}, 3543 (1988);
  Z.\ Phys.\  C {\bf 49} (1991) 601.
\bibitem{an}
  P.~C.~Argyres and C.~R.~Nappi,
  Phys.\ Lett.\  B {\bf 224}, 89 (1989).
\bibitem{s}
  J.~S.~Schwinger,
  Phys.\ Rev.\  {\bf 82}, 664 (1951).
\bibitem{fpo}
  S.~Ferrara and M.~Porrati,
  Mod.\ Phys.\ Lett.\  A {\bf 8}, 2497 (1993)
  [arXiv:hep-th/9306048].
\bibitem{no}
N.~K.~Nielsen and P.~Olesen,
  Nucl.\ Phys.\  B {\bf 144}, 376 (1978).
\bibitem{fpt}
  S.~Ferrara, M.~Porrati and V.~L.~Telegdi,
  Phys.\ Rev.\  D {\bf 46}, 3529 (1992).
\bibitem{pr}
  M.~Porrati and R.~Rahman,
  Nucl.\ Phys.\  B {\bf 814}, 370 (2009)
  [arXiv:0812.4254 [hep-th]].
\end{thebibliography}
\end{document}